\documentclass[prx,amsmath,amssymb,twocolumn]{revtex4}

\usepackage{graphicx}
\usepackage{subfigure}
\usepackage{adjustbox}
\usepackage{bm}
\usepackage{color}
\usepackage{braket}
\usepackage{standalone}
\usepackage{multirow}
\usepackage{tikz}
\usepackage{mathrsfs}
\usepackage{diagbox}
\usepackage{array}
\usepackage{bbold}
\usepackage[colorlinks,bookmarks=true,citecolor=blue,linkcolor=red,urlcolor=blue]{hyperref}

\begin{document}
\title{Dissipative analog of four-dimensional quantum Hall physics}

\author{Fanny Terrier$^{1,2}$ and Flore K. Kunst$^{2, 3}$}

\affiliation{$^1$ Master de Physique Fondamentale, Universit\'{e} Paris-Saclay, 91405 Orsay, France
\\ $^2$ Department of Physics, Stockholm University, AlbaNova University Center, 106 91 Stockholm, Sweden
\\ $^3$ Max-Planck-Institut f\"{u}r Quantenoptik, Hans-Kopfermann-Stra{\ss}e 1, 85748 Garching, Germany}
\date{\today}

\begin{abstract}
Four-dimensional quantum Hall (QH) models usually rely on synthetic dimensions for their simulation in experiment. Here, we study a QH system which features a nontrivial configuration of three-dimensional Weyl cones on its boundaries. We propose a three-dimensional analog of this model in the form of a dissipative Weyl semimetal (WSM) described by a non-Hermitian (NH) Hamiltonian, which in the long-time limit manifests the anomalous boundary physics of the four-dimensional QH model in the bulk spectrum. The topology of the NH WSM is captured by a three-dimensional winding number whose value is directly related to the total chirality of the surviving Weyl nodes. Upon taking open boundary conditions, instead of Fermi arcs, we find exceptional points with an order that scales with system size.
\end{abstract}

\maketitle

\section{Introduction}

Topological phases of matter have been at the forefront of research in condensed matter physics over the last decades \cite{Klitzing1980, Haldane1988, Laughlin1981, Thouless1982, Kane2005, Kane2005a, Bernevig2006, Bernevig2006a}. The communality between models describing such phases is the presence of robust states, which appear in an anomalous configuration on the boundaries of the system and whose existence is described by a topological invariant as dictated by the bulk-boundary correspondence \cite{Hasan2010}. An early example of a topological model is the four-dimensional quantum Hall (QH) model \cite{Zhang2001, Qi2010}, which may feature three-dimensional Weyl cones on its boundaries captured by the second Chern number. While many proposals exist for the experimental simulation of four-dimensional QH systems ranging from photonic setups \cite{Ozawa2016, Lu2018, Zhang2019Entangled} to ultracold atoms \cite{Price2015, Price2016, Price2018} and electrical circuits \cite{Ezawa2019, Lee2018, Yu2019},
only three experiments have been reported \cite{Zilberberg2018, Lohse2018, Wang2020} which make use of topological pumping in two-dimensional setups \cite{Zilberberg2018, Lohse2018} and of electrical circuits \cite{Wang2020}. Additionally, in Ref.~\onlinecite{Sugawa2018}, the second Chern number associated with a five-dimensional Yang monopole is measured through quantum simulation. Due to the physical obstruction of having four spatial dimensions, the majority of the proposals \cite{Ozawa2016, Lu2018, Zhang2019Entangled, Price2015, Price2016, Price2018} as well as the experiments \cite{Zilberberg2018, Lohse2018, Sugawa2018} necessarily rely on synthetic dimensions rendering the physical realization of this system highly non-trivial. Here, we propose an alternative approach to probe the topology of the four-dimensional QH system, namely, through studying a topologically equivalent, three-dimensional analog with dissipation in the form of a \emph{non-Hermitian Weyl semimetal}.

Weyl semimetals (WSMs) are three-dimensional topological models, which feature Fermi arcs on the surfaces that connect Weyl cones in the bulk \cite{Volovik2003, Murakami2007, Wan2011, Burkov2011, Xu2015, Lu2015, Lv2015}. As such, a WSM can be seen as the surface realization of the four-dimensional QH model with one important caveat: The total chirality of the Weyl cones appearing in the bulk spectrum has to disappear, meaning that an anomalous configuration of Weyl cones cannot exist \cite{Nielsen1981}. However, if we allow for the system to be dissipative, this obstruction can be circumvented. Indeed, we find that introducing modulated gain and loss in the WSM, which renders the Hamiltonian of the model non-Hermitian (NH) (see Ref.~\onlinecite{Bergholtz2019} for a recent review),
results in the appearance of an anomalous configuration of Weyl cones in the bulk spectrum in the \emph{long-time limit}: The spectrum of NH Hamiltonians is generally complex, where the imaginary part is associated with the inverse lifetime, such that only states with $\textrm{Im}(E)>0$ survive when $t \rightarrow \infty$ \cite{Lee2019}.

The three-dimensional NH WSM studied in this paper is ensured to be equivalent to the four-dimensional QH system by making use of arguments that are grounded in topology. Specifically, this means that there exists a connection between the topological invariants of the two models: Whereas the chiralities of the Weyl cones that appear in the bulk spectrum of the NH WSM are found to be directly related to the values of the second Chern number of the QH model, we also find that the total chirality of the Weyl cones in the positive imaginary energy plane is given by a three-dimensional \emph{spectral} winding number \cite{Gong2018}, which captures the point-gap topology of the NH WSM, where point gaps are a purely NH phenomenon with the complex energy bands not crossing a reference point \cite{Gong2018} [cf. Fig.~\ref{fig:3d_nh_pbc}(a)]. Our three-dimensional NH WSM thus indeed mimics the four-dimensional QH model.

In addition to studying the bulk properties of the NH WSM, we also investigate its boundary behavior. Interestingly, instead of finding Fermi arcs, we find that the boundaries feature exceptional points (EPs) with an order scaling with system size, where
EPs are degeneracies in the complex spectrum at which both the eigenvalues and the eigenvectors coalesce, thus rendering the Hamiltonian matrix defective at these points, where
the order of an EP is set by the number of eigenvectors that collapse \cite{Heiss2012}.
At the high-order EPs in the model studied here, all bulk states are found to collapse and localize to either one of the boundaries.
The appearance of these EPs is in line with the recent proposal that a nontrivial point-gap topology is linked to the NH skin effect \cite{Borgnia2020, Okuma2019, Zhang2019},
which refers to the piling up of bulk states at the boundaries,
and the NH skin effect is in turn linked to the appearance of or vicinity to EPs of this kind \cite{Xiong2018, Kunst2019}. We nevertheless do not observe a piling up of bulk states away from the EPs, which may be attributed to the fact that the EPs are well isolated from the rest of the spectrum.

This paper is organized as follows: In Sect.~\ref{sec:4D_QHE}, we briefly summarize the four-dimensional QH system, which is followed by the derivation of the NH WSM model and a discussions of its properties in Sect.~\ref{sec:3D_NH_WSM}. We conclude with a discussion in Sec.~\ref{sec:discussion}.

\section{Four-dimensional quantum Hall model} \label{sec:4D_QHE}

We study the following Bloch Hamiltonian, which describes a four-dimensional QH system:
\begin{equation}
    H_\textrm{QH}({\bf k}) = {\bf d} ({\bf k}) \cdot \boldsymbol\Gamma, \label{eq:4D_dirac_Ham}
\end{equation}
where ${\bf k} \equiv (k_x, k_y, k_z, k_w)$,
\begin{equation}
    \mathbf{d}({\bf k}) = \left(
    h + \sum_{i} \cos k_i , \,
    \sin k_x, \,
    \sin k_y, \, 
    \sin k_z, \,
    \sin k_w \right) \label{eq:4D_dirac_Ham_d_vec}
\end{equation}
with $i \in \{x,y,z,w\}$, and $\mathbf{\Gamma} = (\Gamma_1,\Gamma_2,\Gamma_3,\Gamma_4,\Gamma_5)$ is the vector of gamma matrices respecting the Clifford algebra with $\Gamma_1= - \sigma_2 \otimes \sigma_0$, $\Gamma_2 = \sigma_1 \otimes \sigma_1$, $\Gamma_3 = \sigma_1 \otimes \sigma_2$, $\Gamma_4 = \sigma_1 \otimes \sigma_3$, and $\Gamma_5 = \sigma_3 \otimes \sigma_0$, with $\sigma_0$ the identity and $\sigma_j$ the Pauli matrices \cite{Qi2010}. The eigenvalues for Dirac Hamiltonians of the type of Eq.~(\ref{eq:4D_dirac_Ham}) can be straightforwardly computed, and read
\begin{equation}
E_{\textrm{QH}, \pm} ({\bf k}) = \pm |{\bf d} ({\bf k})|, \label{eq:eigenvalues_4d_QHI}
\end{equation}
where both $E_{\textrm{QH}, +}({\bf k})$ and $E_{\textrm{QH}, -}({\bf k})$ are doubly degenerate. Gap closings corresponding to four-dimensional massless Dirac cones occur when $E_{\textrm{QH}, +} ({\bf k}) = E_{\textrm{QH}, -} ({\bf k}) = 0$, i.e., when
\begin{align*}
{\bf k} &= (0,0,0,0), & h&= -4,\\
{\bf k} &= P[(\pi,0,0,0)],  & h &= -2, \\
{\bf k} &= P[(\pi,\pi,0,0)],  & h &= 0, \\
{\bf k} &= P[(\pi,\pi,\pi,0)],  & h &= 2, \\
{\bf k} &= (\pi, \pi, \pi, \pi), & h &= 4,
\end{align*}
with $P[(\ldots)]$ denoting all perturbations of the coordinates.

\begin{figure}[t]
    \centering
    \includegraphics[width=\linewidth]{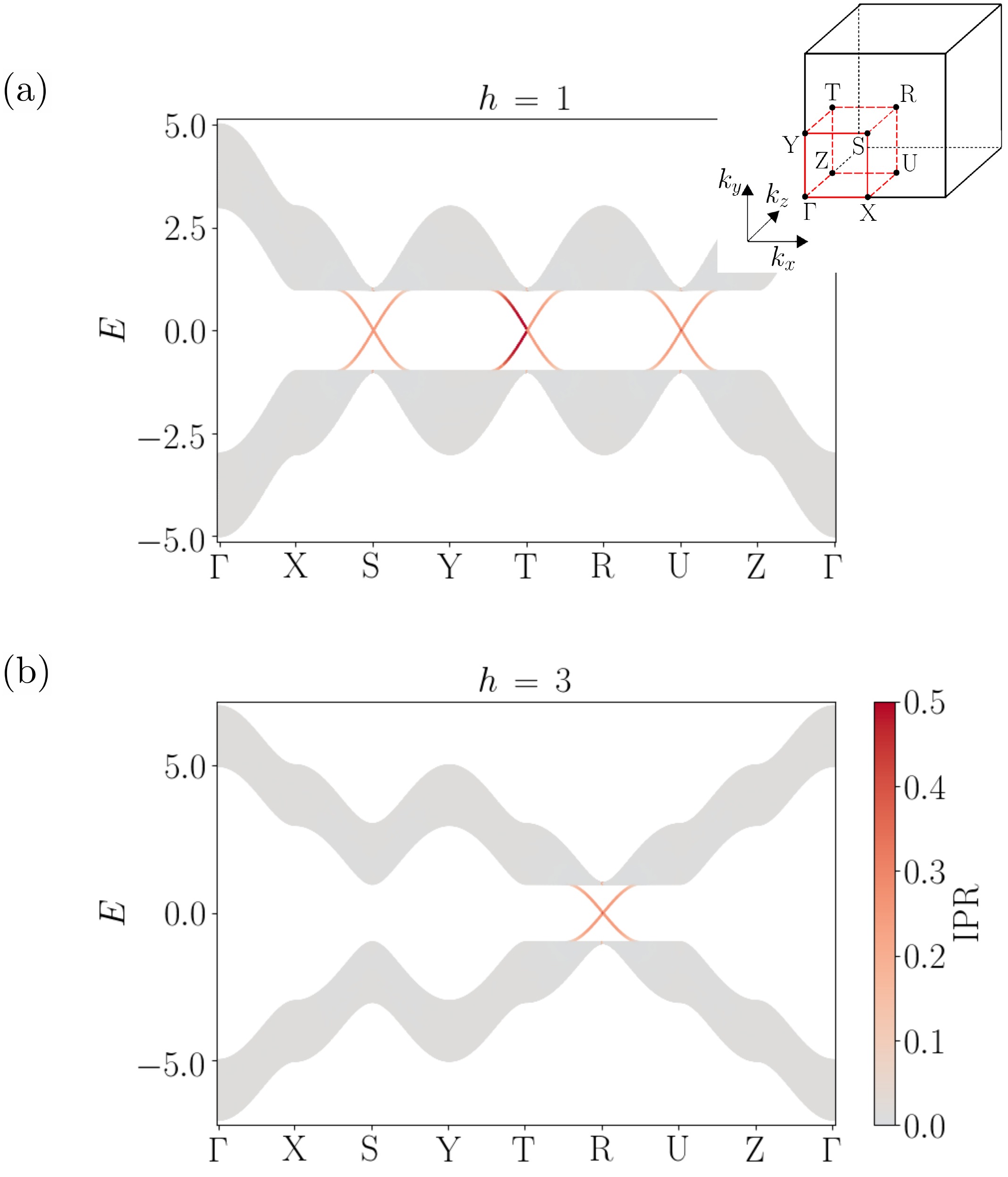}
    \caption{Band spectrum of the four-dimensional QH system with OBC in $w$ with $N = 50$ unit cells for (a) $h=1$ and (b) $h=3$ along a path in the three-dimensional surface BZ, which is shown in the inset in (a). In agreement with the Bloch spectrum, this spectrum is doubly degenerate such that each band-gap crossing corresponds to a doubly degenerate Weyl cone. The color of the bands with the color bar shown in (b) corresponds to the inverse participation ratio (IPR), $I_n = \sum_{j=1}^{4N} |\Psi_{n,j}|^4/(\sum_{j=1}^{4N}|\Psi_{n,j}|^2)^2$, where $\Psi_{n,j}$ is the eigenfunction associated with energy band $E_n$, $j$ is the site index, and there are a total of $4N$ sites. The IPR measures the localization of the eigenstates: It goes to zero for extended states, whereas it acquires a finite value for localized states.}
    \label{fig:4d_qhe_open_bc}
\end{figure}

The topological invariant associated with this model is the second Chern number $C_2$,
\begin{equation}
    C_2 = \frac{3}{8 \pi^2} \int_\textrm{BZ} \epsilon^{abcde} \hat{d}_a \partial_{k_x}\hat{d}_b \partial_{k_y}\hat{d}_c \partial_{k_z}\hat{d}_d \partial_{k_w}\hat{d}_e, \label{eq:definition_second_chern_number}
\end{equation}
where the integral is taken over the Brillouin zone (BZ), $\epsilon^{abcde}$ is the antisymmetric Levi-Civita symbol, $a,b,c,d,e \in \{1,2,3,4,5\}$, $\hat{d_i} \equiv d_i/|{\bf d}|$, and leads to \cite{Qi2010}
\begin{equation}
    C_2 = 
  \begin{cases} 
  0, & \quad |h|>4, \\ 
  1, & \quad -4 < h < -2, \\
  -3, & \quad -2<h<0,\\
  3, & \quad 0<h<2,\\
  -1, & \quad 2< h <4.
  \end{cases}  \label{eq:results_second_Chern_number}
\end{equation}
The value of the second Chern number $C_2$ is related to the presence of three-dimensional Weyl cones on the boundaries of the four-dimensional lattice model, where the number of boundary Weyl cones corresponds to $|C_2|$, while the chirality of the cones is given by $\textrm{sign}(C_2)$ \cite{Qi2010}. Indeed, when we take open boundary conditions (OBCs), we find Weyl cones at the high-symmetry points of the three-dimensional boundary BZ, as shown in Fig.~\ref{fig:4d_qhe_open_bc}. Unlike the two-dimensional QH system, four-dimensional QH physics may be realized in models with time-reversal symmetry \cite{Bernevig2013}, and indeed we find that $T H_\textrm{QH}^*(-{\bf k}) T^{-1} = H_\textrm{QH}({\bf k})$ with $T = i \Gamma_3$ and $T T^* = -1$. The model in Eq.~(\ref{eq:4D_dirac_Ham}) thus belongs to symmetry class AII \cite{Altland1997, Schnyder2008, Kitaev2009, Ryu2010}.

\section{Non-Hermitian Weyl semimetal} \label{sec:3D_NH_WSM}

Here we introduce a three-dimensional NH model, which is topologically equivalent to the four-dimensional QH model and realizes Weyl cones with a nonzero total chirality in the positive imaginary plane of the spectrum. We start by reexpressing the Hamiltonian in Eq.~(\ref{eq:4D_dirac_Ham}) as
\begin{align*}
    H_\textrm{QH}({\bf k}) &= \sum_{i=1}^4 d_i ({\bf k})\Gamma_i + \sin k_w \Gamma_5 \\
    & \equiv \bar{H}({\bf k}) + \sin k_w \Gamma_5,
\end{align*}
where $d_i ({\bf k})$ are the components of the vector ${\bf d}({\bf k})$ defined in Eq.~(\ref{eq:4D_dirac_Ham_d_vec}). $\bar{H}({\bf k})$ has a chiral structure, i.e.,
\begin{align*}
\bar{H} ({\bf k})&= \begin{pmatrix}
0 & i h_0({\bf k}) \sigma_0 + {\bf h}({\bf k}) \cdot \boldsymbol\sigma \\
-i h_0({\bf k}) \sigma_0 + {\bf h}({\bf k}) \cdot \boldsymbol\sigma & 0
\end{pmatrix} \\
& \equiv \begin{pmatrix}
0 & H_0 ({\bf k}) \\
H_0^\dagger ({\bf k}) & 0 
\end{pmatrix},
\end{align*}
where $\boldsymbol\sigma$ is the vector of Pauli matrices, $h_0({\bf k}) \equiv d_1({\bf k})$ and ${\bf h}({\bf k}) \equiv \left(d_2({\bf k}), d_3({\bf k}), d_4({\bf k}) \right)$, and anticommutes with $\Gamma_5$, i.e., $\{\bar{H} ({\bf k}) , \Gamma_5\} = 0$.
The term proportional to $\Gamma_5$ can thus be interpreted as a mass term.
Therefore,
the gap in the energy spectrum of $H_\textrm{QH} ({\bf k})$ closes when there is a gap closing in the spectra of $\bar{H} ({\bf k})$ and $\sin k_w \Gamma_5$ simultaneously. This means that the topology of $H_\textrm{QH} ({\bf k})$ is captured by $\bar{H} ({\bf k}, k_w = 0)$ and $\bar{H} ({\bf k}, k_w = \pi)$.
Due to the chiral symmetry (given by $\Gamma_5$) of $\bar{H} ({\bf k}, k_w = 0, \pi)$, we can compute a winding number to characterize the topological properties of these Hamiltonians \cite{Schnyder2008, Ryu2010}.
This means that we need to compute the winding number for one of the NH off-diagonal components of $\bar{H} ({\bf k}, k_w = 0, \pi)$, i.e., $H_0({\bf k}, k_w=0)$, $H_0({\bf k}, k_w=\pi)$, $H_0^\dagger({\bf k}, k_w=0)$, or $H_0^\dagger({\bf k}, k_w=\pi)$. We find that equivalent results can be derived for all of these four NH choices:
The term ${\bf h}({\bf k}) \cdot \boldsymbol\sigma$, which determines the gap closings of these NH Hamiltonians, is Hermitian and independent of $k_w$, which means that all four possible choices of NH Hamiltonians are topologically equivalent. Indeed, the only difference between the four Hamiltonians is a constant and/or overall minus sign that goes with the identity term. Therefore, we may focus on the topological characterization of the following NH Hamiltonian:
\begin{align}
H_\textrm{NH}({\bf k}) &= i \left(h + \sum_{i \in \{x,y,z\}}  \cos k_i \right) \sigma_0 + {\bf h}({\bf k}) \cdot \boldsymbol\sigma, \label{eq:NH_3d_ham} \\
{\bf h}({\bf k}) & = \left(\sin k_x, \, \sin k_y \, \sin k_z \right), \nonumber
\end{align}
which is topologically equivalent to $H_0^{(\dagger)}({\bf k}, k_w=0, \pi)$. This model is time-reversal symmetric according to $H_\textrm{NH}({\bf k}) = T H_\textrm{NH}^T(-{\bf k}) T^{-1}$ with $T = i \sigma_y$ and $T T^* = -1$, and as such belongs to class AII$^\dagger$ \cite{Okuma2019, Kawabata2018Topology}. This is in line with the statement in Ref.~\onlinecite{Lee2019} that a $(d+1)$-dimensional Hermitian model in class $s$ is characterized by a $d$-dimensional NH model in class $s^\dagger$.
The topological properties of the four-dimensional QH model in Eqs.~(\ref{eq:4D_dirac_Ham}) and (\ref{eq:4D_dirac_Ham_d_vec}) are thus captured by the three-dimensional NH model in Eq.~(\ref{eq:NH_3d_ham}).

The eigenvalues of $H_\textrm{NH}({\bf k})$ are $E_{\pm} ({\bf k}) = i (h + \sum_i \cos k_i) \pm |{\bf h}({\bf k})|$, such that gap closings, $E_{+} ({\bf k}) = E_{-} ({\bf k})$, appear when $|{\bf h}({\bf k})| = 0$ is satisfied, i.e., when ${\bf k} = (0,0,0)$, $P[(\pi,0,0)]$, $P[(\pi,\pi,0)]$, and $(\pi, \pi, \pi)$. Despite the abundance of second-order exceptional lines in the complex spectra of three-dimensional NH models \cite{Berry2004, Budich2019, Yoshida2019}, these eight gap closings correspond to ``ordinary" three-dimensional Weyl cones and are degenerate in the usual sense, i.e., the geometric and algebraic multiplicities of the gap closings are equivalent. Indeed, the vector ${\bf h}({\bf k})$, which determines the occurrence as well as the nature of the degeneracies, is entirely real, i.e., ${\bf h}({\bf k}) \in \mathbb{R}^3$, and as such prohibits the appearance of exceptional structures in the complex spectrum of $H_\textrm{NH}({\bf k})$. The Hamiltonian $H_\textrm{NH}({\bf k})$ thus describes a three-dimensional NH Weyl semimetal.

\begin{figure}[t]
    \centering
    \includegraphics[width=0.95\linewidth]{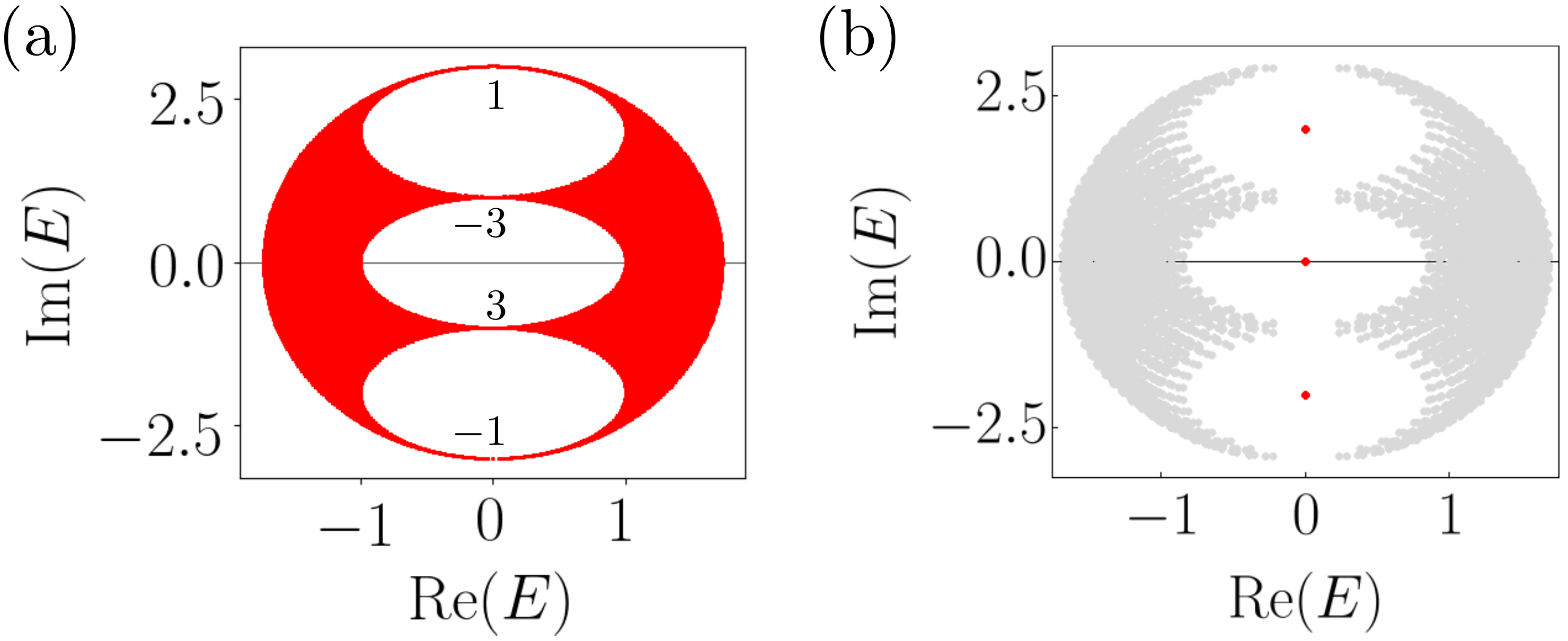}
    \caption{Spectrum of the three-dimensional NH Hamiltonian in Eq.~(\ref{eq:NH_3d_ham}) for (a) periodic boundary conditions (PBCs) and (b) OBCs with $h = 0$. (a) The chirality $\chi$ of the gap closings is explicitly indicated in agreement with Eq.~(\ref{eq:3d_NH_gap_closings}), and the winding number $w_{3D}$ equals $2$ for this plot ($h=0$). (b) Spectrum for OBC in $x$ for $N=14$ unit cells and with $k_y$ and $k_z$ parametrized between $-\pi$ and $\pi$ with steps of $2 \pi / 40$. The red points appearing inside the gap are EPs with an algebraic multiplicity of $2N$ and a geometric multiplicity of 2. The EPs have energies $E = -2i$, $0$ and $2i$ when $(k_y, k_z) = (\pi,\pi)$, $P[(0, \pi)]$, and $(0,0)$, respectively. This spectrum is computed analytically due to the strong numerical instability existing at the EPs (see Appendix~\ref{app:num_instability_EPs} for more details).}
    \label{fig:3d_nh_pbc}
\end{figure}

Due to the conventional nature of the gap closings, we can ascribe a chirality $\chi$ to each Weyl cone in a similar fashion as in the Hermitian case: By performing a low-energy approximation around the Weyl points, we obtain an effective Hamiltonian of the form $M {\bf k} \cdot \boldsymbol\sigma$ with ${\bf k} = (k_x, k_y, k_z)$, where the determinant of the matrix $M$ determines the chirality, $\textrm{det}M = \chi$. This leads to the following energies and chiralities at the eight degeneracies:
\begin{align}
&{\bf k} = (0,0,0), & E &= i (h + 3), & \chi &= 1,\nonumber \\
&{\bf k} = P[(\pi,0,0)], & E &= i(h +1), & \chi &= -1, \nonumber \\
&{\bf k} = P[(\pi,\pi,0)], & E &= i (h - 1) , & \chi &= 1,\nonumber \\
&{\bf k} = (\pi, \pi, \pi), & E &= i(h - 3), & \chi &= -1. \label{eq:3d_NH_gap_closings}
\end{align}
We observe that the three Weyl cones at ${\bf k} = P[(\pi,0,0)]$ and at ${\bf k} = P[(\pi,\pi,0)]$ have the same energy, such that we find four band-gap closings in the spectrum with chiralities $1$, $-1$, $3$, and $-3$, as shown in Fig.~\ref{fig:3d_nh_pbc}(a). We notice that these chiralities correspond to the values of the second Chern number $C_2$ of the four-dimensional QH model [cf. Eq.~(\ref{eq:results_second_Chern_number})], which is in line with the expectation that $H_\textrm{NH}({\bf k})$ [cf. Eq.~(\ref{eq:NH_3d_ham})] realizes the boundary behavior of the four-dimensional QH model: Indeed, the gap closings in the spectrum of $H_\textrm{NH}({\bf k})$ correspond to all four different Weyl-cone configurations that can appear on the boundary of the four-dimensional QH model as determined by the second Chern number $C_2$. 

The three-dimensional winding number reads
\begin{equation}
    w_{3D} = \frac{1}{24 \pi^2} \int_{BZ} d^3 k \, \epsilon^{\mu\nu\sigma}Tr\left(Q_{\mu}Q_{\nu}Q_{\sigma}\right), \label{eq:definition_winding_number}
\end{equation}
where the integral is over the BZ, $\epsilon^{\mu\nu\sigma}$ is the Levita-Civita symbol, $\mu, \nu, \sigma \in \{k_x, k_y, k_z\}$, and $Q_{\alpha} \equiv (H_{NH})^{-1} \partial_\alpha H_{NH}$ \cite{Schnyder2008, Ryu2010, Kawabata2018Topology}. Computing $w_{3D}$ for $H_\textrm{NH}({\bf k})$ with $E_B = 0$ as a reference point, we find
\begin{equation*}
    w_{3D} = 
  \begin{cases} 
  0, & \quad |h| > 3, \\ 
  -1, & \quad -3 < h < -1, \\
  2, & \quad -1<h<1,\\
  -1, & \quad 1<h<3.
  \end{cases}  \label{eq:results_three_dim_winding_number}
\end{equation*}
The value of the winding number can be understood from the energy eigenvalues: When $|h|>3$, the energy bands do not wind around $E = 0$ such that $w_{3D} = 0$. When $|h|<3$, on the other hand, the bands wind around $E=0$ and $w_{3D}$ acquires a nonzero value.
In fact, we notice that the value of the winding number corresponds to the difference between the total negative chirality, $\chi_-^{\textrm{Im}E>0}$, and the total positive chirality, $\chi_+^{\textrm{Im}E>0}$, of the cones with positive imaginary energy, i.e., $w_{3D} = \chi_-^{\textrm{Im}E>0} - \chi_+^{\textrm{Im}E>0}$, and as such is indirectly related to the values of $C_2$. For example, when $-1 < h < 1$, there are three cones with negative chirality, $\chi_-^{\textrm{Im}E>0} = 3$, and one cone with positive chirality, $\chi_+^{\textrm{Im}E>0} = 1$, with $\textrm{Im}(E)>0$, such that $w_{3D} = 3 - 1 =2$ in this energy region. This result is in analogy with the result in Ref.~\onlinecite{Lee2019}, where the one-dimensional winding number is found to correspond to the difference between the number of left- and right-moving modes with positive imaginary energy appearing in the one-dimensional Hatano-Nelson model \cite{Hatano1996}. We thus find that the value of the three-dimensional winding number $w_{3D}$ corresponds to the total chirality of the Weyl cones surviving in the long-time limit, such that a nonzero $w_{3D}$ signals the realization of anomalous Hermitian boundary physics in the bulk on the NH model.

As the three-dimensional NH model realizes Weyl cones in the bulk of its spectrum in the same fashion as in the Hermitian case, we naively expect the appearance of Fermi arcs on the surfaces connecting the cones. In Fig.~\ref{fig:3d_nh_pbc}(b), we plot the spectrum for OBC in $x$ and see that, instead of Fermi arcs, EPs (in red) appear, whose orders scale with system size. These EPs appear when $k_y, k_z \in \{0 , \pi\}$ and have an algebraic multiplicity of $2N$, with $N$ the total number of unit cells, while the geometric multiplicity remains $2$ regardless of the system size (see Appendix~\ref{app:analytical_comp_eigensystem} for an analytical derivation of these results). The appearance of these EPs, where all eigenstates collapse onto two eigenstates of which one is localized entirely on one boundary and the other entirely on the other (see Appendix~\ref{app:analytical_comp_eigensystem}), is consistent with the statement in Refs.~\onlinecite{Borgnia2020, Okuma2019, Zhang2019}, where it is shown that having a topologically non-trivial point gap results in the appearance of skin states, i.e., bulk states that are localized to the boundary. However, as the EPs are well isolated from the rest of the bands in the spectrum, the states away from the EPs do not pile up and behave as ordinary bulk states (see Appendix~\ref{app:analytical_comp_eigensystem} for the derivation of the localization of the bulk states), as is also found in Ref.~\onlinecite{Okuma2019}, where a two-dimensional version of our model is studied. This is consistent with the logic presented in Refs.~\onlinecite{Xiong2018, Kunst2019}, where the presence of higher-order EPs is linked to the NH skin effect. Indeed, only when the other states can ``feel" the presence of such EPs, will they also start to pile up at the boundaries. As such, this model provides an interesting insight in the interplay of the presence of EPs with high order and the piling up of bulk states.

\begin{figure}[b]
    \centering
    \includegraphics[width=\linewidth]{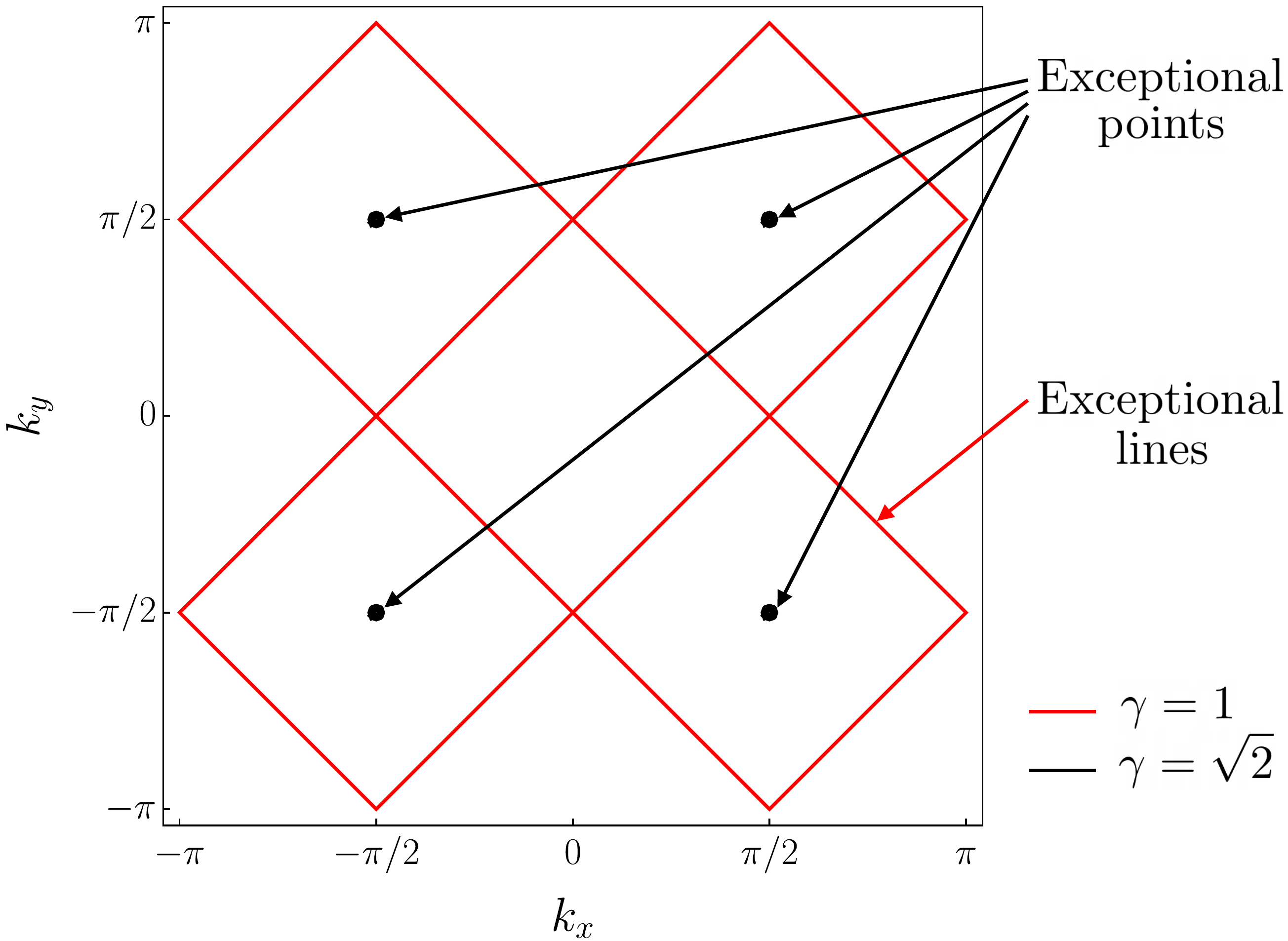}
    \caption{Exceptional lines (red) and points (black) in the bulk spectrum of the model in Eq.~(\ref{eq:NH_3d_ham}) with the addition of $-i \gamma \sigma_3$ for $k_z \in \{0, \pi\}$, and $\gamma = 1$ and $\gamma = \sqrt{2}$, respectively. The exceptional lines have energy $E =  i \left(h + \cos k_x + \cos k_y\right)$ with $k_x$ and $k_y$ solutions to $\sin^2 k_x +\sin^2 k_y = 1$, while the exceptional points all have the energy $E =  i \, h$.}
    \label{fig:exceptional_lines_and_points}
\end{figure}

\section{Discussion} \label{sec:discussion}

In this work, we have studied a three-dimensional analog of a four-dimensional QH model in the form of an NH WSM. The NH model in Eq.~(\ref{eq:NH_3d_ham}) is special in the sense that its gap closings can never be exceptional as ${\bf h}({\bf k}) \in \mathbb{R}^3$ as discussed above. By introducing NH terms into the vector ${\bf h}({\bf k})$, however, it is straightforward to find exceptional structures as shown in Fig.~\ref{fig:exceptional_lines_and_points}.

NH Hamiltonians are well suited to describe classical systems with dissipation (see, e.g., Refs.~\onlinecite{Lu2014, Brandenbourger2019, Helbig2019}), but also find applications in the quantum realm (see, e.g., Refs.~\onlinecite{Carmichael1993, Kozii2017, Yoshida2018, Bergholtz2019nonHermitian}), and it may thus be possible to access the NH model in Eq.~(\ref{eq:NH_3d_ham}) in experiment. One potential platform for the realization of our model is in photonic setups, where gain and loss can be implemented in experiment \cite{Zhao2018, Parto2018, Malzard2018}. In particular, three-dimensional photonic crystals form suitable candidates, where gain can be introduced through external pumping, while the presence of defect states provides a mechanism for radiative losses \cite{Joannopoulos2008}. Here, the real part of the band spectrum corresponds to the resonance frequency of the crystal, whereas the imaginary part sets the linewidth of the resonances \cite{Zhou2018}. In Ref.~\onlinecite{Zhou2018}, the real part of the spectrum of the realization of an NH model in a two-dimensional setup is accessed through measuring the isofrequency contours of the crystal, whereas in Refs.~\onlinecite{Lu2015, Yang2017} it is demonstrated that both the bulk dispersion \cite{Lu2015} and the boundary features \cite{Yang2017} of a Hermitian WSM are experimentally accessible. Therefore, if one were to implement our NH model, we expect that one would be able to resolve the Weyl cones in the bulk, as well as EPs on the boundaries.

Another promising platform for the realization of our model is that of electric circuits \cite{Albert2015, Ningyuan2015}, which were used very recently to realize a four-dimensional quantum Hall model in class AI with $T T^*=1$ \cite{Wang2020} as well as a Hermitian WSM \cite{Lu2019}, while the realization of an NH model has also been reported \cite{Helbig2019}. In these circuits, capacitors and inductors act as Hermitian components, and resistors and amplifiers as anti-Hermitian ones \cite{Helbig2019}. Circuits can be used to realize models of any dimension, and in particular three-dimensional models can be realized by simply stacking two-dimensional circuit boards, as in Refs.~\cite{Wang2020, Lu2019}. Alternatively, our three-dimensional NH model could also be realized in a lower-dimensional circuit as the nodes, which play the role of the lattice sites, can be connected along more than three directions \cite{Lee2018}. The circuits are governed by Kirchhoff's law, which relates the admittance matrix (or Laplacian matrix) $J$ to the current $I$ and voltage $V$ via $I = J V$, where the admittance matrix $J/(i \omega)$ plays the role of the Hamiltonian, with $\omega$ the current frequency \cite{Ningyuan2015}. It is thus possible to realize non-interacting Hamiltonians such as the one in Eq.~(\ref{eq:NH_3d_ham}) through appropriately arranging the various electronic tools that are available. Moreover, by changing the connectivity between the nodes, one can interpolate between PBC and OBC, such that both of these cases can be studied.
The eigenvalues of the system can subsequently be accessed by measuring the impedance $Z$, which is related to the admittance matrix via its inverse and both its real and imaginary parts are readily accessible in experiment \cite{Lee2018}.
Indeed, in Ref.~\onlinecite{Helbig2019}, this technique was used to measure the real and imaginary parts of the eigenspectrum of an NH anisotropic chain. Additionally, it is shown in that work \cite{Helbig2019} that the NH skin effect is manifested as a nonlocal voltage response in the circuit. We thus expect that both the PBC and OBC spectrum of our three-dimensional model should be directly measurable via the impedance, and that it is also possible to study the skin effect when performing response measurements when the system is operating at one of the EPs in the OBC system.

A connection between $(d+1)$-dimensional Hermitian topological insulators and $d$-dimensional NH models with nontrivial point-gap topology was recently also made on an abstract level by Lee et al. in Ref.~\onlinecite{Lee2019}. As an illustration, they present one- and two-dimensional NH examples, which emulate the edge and surface physics of a two-dimensional Chern insulator and three-dimensional chiral topological insulator, respectively, which is complemented by the higher-dimensional model treated in this work. A three-dimensional model mimicking four-dimensional QH physics was also proposed in a Floquet system in Ref.~\onlinecite{Sun2018}, where it is shown that it is possible to find Weyl cones with a nonzero total chirality by making use of the adiabatic limit. We point out that the approach used in that work \cite{Sun2018} is different from the one in this paper.

We note that the exotic feature displayed by the NH WSM under OBC, namely, the appearance of EPs with an order scaling with system size, has previously been associated with the breaking of bulk-boundary correspondence \cite{Xiong2018, Kunst2019}, as well as an extreme spectral instability against boundary conditions \cite{Krause1994, Xiong2018, Lee2016, Kunst2018, Koch2019, Herviou2019, Kunst2019}. It may thus be tempting to attribute the same properties to the model in this work. The models studied in those papers, however, all have so-called line gaps, where line gaps are a straightforward generalization of the gap in Hermitian spectra with the bands not crossing a line \cite{Gong2018}, whereas the model in Eq.~(\ref{eq:NH_3d_ham}) only features point gaps. Indeed, comparing Figs.~\ref{fig:3d_nh_pbc}(a) and \ref{fig:3d_nh_pbc}(b), we see that there is no spectral instability, and we may thus conclude that these phenomena---a breaking of bulk-boundary correspondence, the NH skin effect, and associated higher-order EPs, and the spectral instability---do not appear as a triad for models with only point gaps in the spectrum.

Lastly, we point out that due to the pivotal role played by time in obtaining an anomalous Weyl-cone configuration in the bulk spectrum of our model, strictly speaking, we need four dimensions, namely three spatial and one time dimension, to realize our NH analog of the four-dimensional QH model. Therefore, while it is thus possible to mimic the boundary physics of $(d+1)$-dimensional Hermitian models in $d$-dimensional NH models, where the dimensions refer to spatial dimensions, one crucially needs $d+1$ spacetime dimensions to obtain the desired bulk behavior.

\acknowledgments{
We thank Emil J. Bergholtz, Jan Carl Budich, and Lo\"{i}c Herviou for useful conversations. We also thank Patrick Emonts for assisting in obtaining analytical results for the OBC spectrum
for larger system sizes.
F.K.K. was funded by the Swedish Research Council (VR) and the Knut and Alice Wallenberg Foundation. F.K.K. was also supported by the Max Planck Institute of Quantum Optics (MPQ) and the Max-Planck-Harvard Research Center for Quantum Optics (MPHQ).
} 

\appendix

\section{Numerical instability at exceptional points} \label{app:num_instability_EPs}

\begin{figure}[t]
    \centering
    \includegraphics[width=0.9\linewidth]{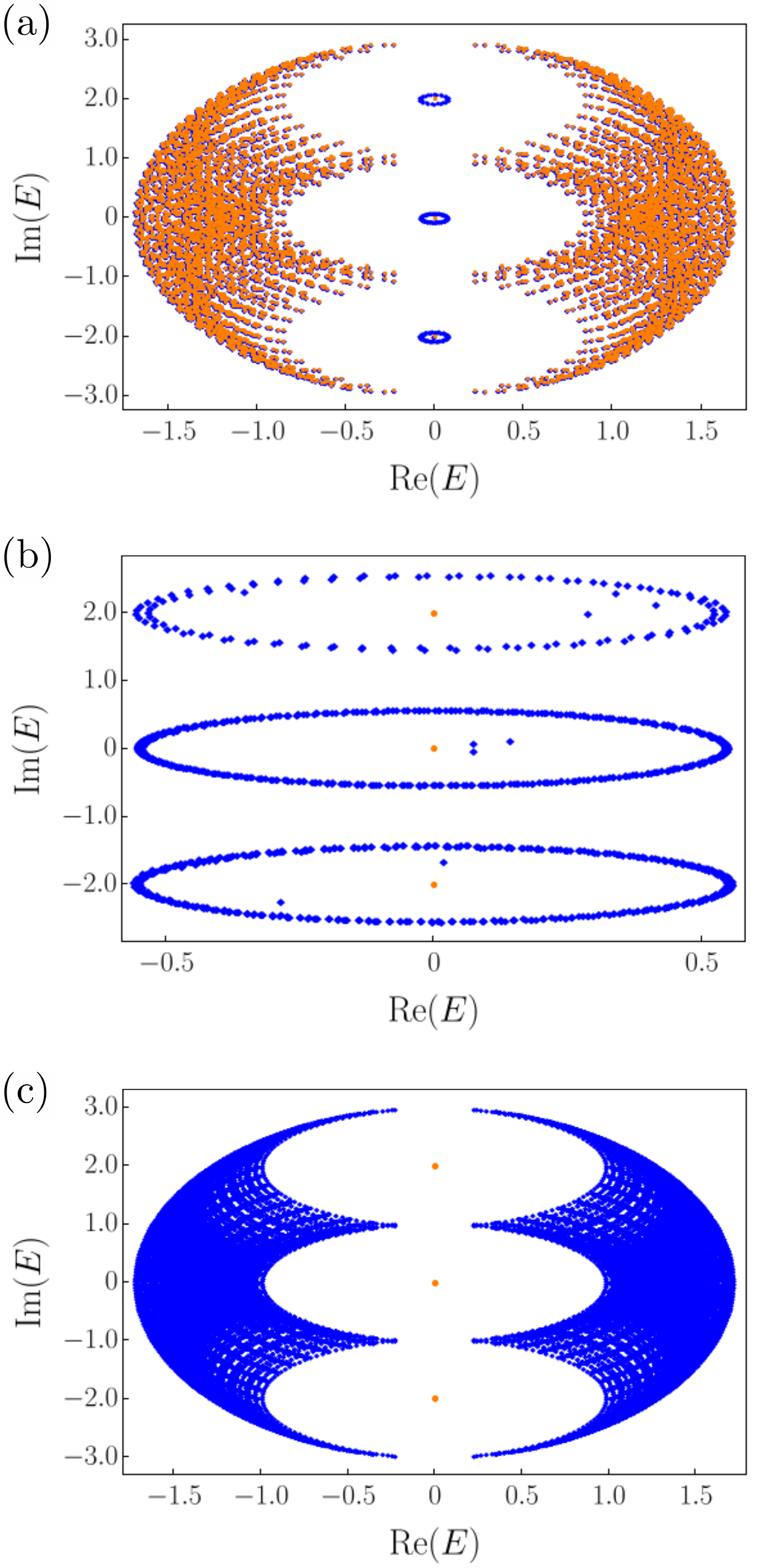}
    \caption{The spectrum of the three-dimensional NH WSM Hamiltonian with OBC in $x$ for $h=0$ with (a) $N=14$ unit cells, (b) the eigenvalues with $k_y, k_z \in \{0, \pi\}$ and $N=60$ unit cells, and (c) $N = 60$ unit cells. The analytically (numerically) computed spectrum is shown in orange (blue).}
    \label{fig:3d_nh_wsm_obc_x_num_vs_an}
\end{figure}

Here we compare the analytically and numerically computed spectra of the three-dimensional NH WSM Hamiltonian with OBC in $x$. In Fig.~\ref{fig:3d_nh_wsm_obc_x_num_vs_an}(a), we plot the OBC spectrum with $N=14$ unit cells with the analytical results in orange and the numerical results in blue, such that the orange (analytical) spectrum is equivalent to the spectrum in Fig.~\ref{fig:3d_nh_pbc}(b). Whereas the results for the bulk spectrum are in good agreement with each other up to machine precision---the difference of the real part of the numerically and analytically obtained eigenvalues is of the order $ \sim 10^{-15}$, whereas the difference of the imaginary part of the numerical and analytical results ranges from $ \sim 10^{-15}$ for most of the eigenvalues to $ \sim 10^{-4}$---the results for the in-gap states, i.e., the EPs, deviate significantly, hinting at a numerical instability at these points. This numerical instability becomes even larger upon increasing the system size as shown in Fig.~\ref{fig:3d_nh_wsm_obc_x_num_vs_an}(b) for $N=60$ unit cells, where the deviation (in blue) away from the EPs (orange) is significant. We point out that the numerical eigenvalues are obtained using the standard eigensolver in Mathematica, and that changing the eigensolver, decreasing the tolerance, and increasing the maximum iteration do not alter the numerical results. Indeed, numerical computations performed in Python ($\textunderscore$geev LAPACK routines) yield the same results. This failure of standard numerical techniques to correctly find the spectrum is due to the extreme defectiveness of the matrix Hamiltonian at the EPs, i.e., there are only two eigenvectors instead of $2N$, which prevents the convergence of numerical values. Due to the apparent correspondence of the analytically and numerically computed bulk bands [cf. Fig.~\ref{fig:3d_nh_wsm_obc_x_num_vs_an}(a)], we plot the bulk bands numerically (in blue) and the EPs analytically (orange) in Fig.~\ref{fig:3d_nh_wsm_obc_x_num_vs_an}(c) for a large system ($N=60$), and see that the bulk gap remains open with the increase in system size. We stress that the results in Fig.~\ref{fig:3d_nh_wsm_obc_x_num_vs_an} tell a cautionary tale, and one needs to be extremely careful when performing numerical computations on NH matrices that have a large defectiveness. Indeed, for such matrices, analytical checks should be made.

\section{Analytical computation of eigensystem} \label{app:analytical_comp_eigensystem}

In this appendix, we compute eigenvalues and eigenvectors for the NH WSM with OBC in $x$.

{\em Eigenvalues.} The Hamiltonian with OBC in $x$ and a total of $N$ unit cells reads
\begin{equation*}
H_{N,x} = {\bf c}^\dagger_N \mathcal{H}_{N,x} {\bf c}_N,
\end{equation*}
where ${\bf c}_N = (c_{1,A}, c_{1,B}, c_{2,A} ,c_{2,B}, \ldots, c_{N,A}, c_{N,B})^T$ with $A, B$ the degrees of freedom and $c_{x,j}^\dagger$ ($c_{x,j}$) creating (annihilating) a state with degree of freedom $j$ in unit cell $x$, such that we find
\begin{equation*}
\mathcal{H}_{N,x} = \begin{pmatrix}
\mathcal{H}_{1,x} & \mathcal{H}_{\perp,x} & 0 & \cdots & 0 \\
\tilde{\mathcal{H}}_{\perp,x} & \mathcal{H}_{1,x} & \mathcal{H}_{\perp,x} & \cdots & 0 \\
0 & \tilde{\mathcal{H}}_{\perp,x} & \mathcal{H}_{1,x} & \cdots & 0 \\
\vdots & \vdots & \vdots & \ddots & \mathcal{H}_{\perp,x} \\
0 & 0 & 0 & \tilde{\mathcal{H}}_{\perp,x} & \mathcal{H}_{1,x}
\end{pmatrix},
\end{equation*}
where
\begin{align*}
\mathcal{H}_{1,x} &= i h_0 \sigma_0 + h_2 \sigma_2 + h_3 \sigma_3, \\
 \mathcal{H}_{\perp,x} &= \frac{i}{2} \left(\sigma_0 + \sigma_1 \right), \qquad  \tilde{\mathcal{H}}_{\perp,x} = \frac{i}{2} \left(\sigma_0 - \sigma_1 \right),
\end{align*}
with $h_0 = h + \cos k_y + \cos k_z$, and $h_2$ and $h_3$ defined in Eq.~(\ref{eq:NH_3d_ham}), i.e., $h_2 = \sin k_y$ and $h_3 = \sin k_z$. The eigenvalues $\lambda$ for this Hamiltonian are found by solving the characteristic equation
\begin{equation*}
\textrm{det} \left(\mathcal{H}_{N,x} - \lambda \mathbb{1}_{2N}\right) = \textrm{det} \, \tilde{\mathcal{H}}_{N,x} = 0,
\end{equation*}
where we define $\tilde{\mathcal{H}}_{N,x} \equiv \mathcal{H}_{N,x} - \lambda \mathbb{1}_{2N}$.
We use Schur's determinant identity
\begin{equation}
\textrm{det}\begin{pmatrix}
A & B \\
C & D
\end{pmatrix} = \textrm{det}D \times \textrm{det}\left(A - B D^{-1} C \right), \label{eq:schurs_determinant_identity}
\end{equation}
to reduce this eigenvalue equation to
\begin{align}
&\textrm{det}\, \tilde{\mathcal{H}}_{1,x} \nonumber \\
& \times \textrm{det}\left(\tilde{\mathcal{H}}_{N-1,x} - \mathcal{H}_{\perp,x}^{N-1 \times 1} \, \tilde{\mathcal{H}}_{1,x}^{-1} \, \tilde{\mathcal{H}}_{\perp,x}^{1 \times N-1} \right)  = 0, \label{eq:eigs_from_det}
\end{align}
where we identify $A = \tilde{\mathcal{H}}_{N-1,x}$, $D = \tilde{\mathcal{H}}_{1,x} $, and 
\begin{align*}
B &= \mathcal{H}_{\perp,x}^{N-1 \times 1} \equiv \begin{pmatrix}
0 \\
\vdots \\
\mathcal{H}_{\perp,x}
\end{pmatrix}, \\
C &= \tilde{\mathcal{H}}_{\perp,x}^{1 \times N-1} = \begin{pmatrix}
0 & \cdots & \tilde{\mathcal{H}}_{\perp,x}
\end{pmatrix},
\end{align*}
such that
\begin{gather*}
\tilde{\mathcal{H}}^{N-1}_{\perp,x} \equiv \mathcal{H}_{\perp,x}^{N-1 \times 1} \, \tilde{\mathcal{H}}_{1,x}^{-1} \, \tilde{\mathcal{H}}_{\perp,x}^{1 \times N-1} \\
= \begin{pmatrix}
0 & \cdots & 0 \\
\vdots & \ddots & \vdots \\
0 & \cdots & \mathcal{H}_{\perp,x} \, \tilde{\mathcal{H}}_{1,x}^{-1} \, \tilde{\mathcal{H}}_{\perp,x}
\end{pmatrix}.
\end{gather*}
Making use of the specific form of the matrices, we find
\begin{equation}
\mathcal{H}_{\perp,x} \, \tilde{\mathcal{H}}_{1,x}^{-1} \, \tilde{\mathcal{H}}_{\perp,x} = \frac{1}{2} \frac{\left(h_2 - i h_3 \right) \sigma_2 + \left(h_3 + i h_2 \right) \sigma_3}{\textrm{det}\, \tilde{\mathcal{H}}_{1,x}}, \label{eq:rescaling_term}
\end{equation}
and
\begin{equation}
\textrm{det}\, \tilde{\mathcal{H}}_{1,x} = \left(i h_0 - \lambda\right)^2 - h_2^2 - h_3^2. \label{eq:det_of_small_block}
\end{equation}
We note that the matrix $\tilde{\mathcal{H}}^{N-1}_{\perp,x}$ has nonzero entries only in its lower right $2 \times 2$ block, such that subtracting it from $\tilde{\mathcal{H}}_{N-1,x}$ only changes the lower right block, i.e.,
\begin{gather*}
\tilde{\mathcal{H}}_{N-1,x} - \tilde{\mathcal{H}}^{N-1}_{\perp,x} = \begin{pmatrix}
\tilde{\mathcal{H}}_{1,x} & \mathcal{H}_{\perp,x} & \cdots & 0 \\
\tilde{\mathcal{H}}_{\perp,x} & \tilde{\mathcal{H}}_{1,x} & \cdots & 0 \\
\vdots & \vdots & \ddots & \mathcal{H}_{\perp,x} \\
0 & 0 & \tilde{\mathcal{H}}_{\perp,x} & \tilde{\tilde{\mathcal{H}}}_{1,x}
\end{pmatrix},
\end{gather*}
where
\begin{equation*}
\tilde{\tilde{\mathcal{H}}}_{1,x} \equiv \tilde{\mathcal{H}}_{1,x} - \mathcal{H}_{\perp,x} \, \tilde{\mathcal{H}}_{1,x}^{-1} \, \tilde{\mathcal{H}}_{\perp,x}.
\end{equation*}
Again applying Schur's determinant identity [cf. Eq.~(\ref{eq:schurs_determinant_identity})], we can reduce Eq.~(\ref{eq:eigs_from_det}) further
\begin{align*}
\textrm{det}\, \tilde{\mathcal{H}}_{N,x} &= \textrm{det}\, \tilde{\mathcal{H}}_{1,x} \times \textrm{det}\, \tilde{\tilde{\mathcal{H}}}_{1,x} \\
& \times \textrm{det}\left(\tilde{\mathcal{H}}_{N-2,x} - \mathcal{H}_{\perp,x}^{N-2 \times 1} \, \tilde{\tilde{\mathcal{H}}}_{1,x}^{-1} \, \tilde{\mathcal{H}}_{\perp,x}^{1 \times N-2} \right)  = 0,
\end{align*}
where $\mathcal{H}_{\perp,x}^{N-2 \times 1} \, \tilde{\tilde{\mathcal{H}}}_{1,x}^{-1} \, \tilde{\mathcal{H}}_{\perp,x}^{1 \times N-2}$ is again only non-zero in the bottom right corner. Repeating this step $N$ times, and defining
\begin{equation*}
a_1 \equiv \tilde{\mathcal{H}}_{1,x}, \qquad a_i \equiv a_1 - \mathcal{H}_{\perp,x} \, a_{i-1}^{-1} \, \tilde{\mathcal{H}}_{\perp,x}, \quad i\geq 2
\end{equation*}
we finally arrive at the expression
\begin{equation}
\textrm{det}\, \tilde{\mathcal{H}}_{N,x} = \prod_i^N \textrm{det} \, a_i = 0. \label{eq:eigenvalue_eq_obc}
\end{equation}
We thus find that the determinant of the $2N$-dimensional OBC Hamiltonian can be expressed in terms of a product of $N$ determinants of two-dimensional matrices. We note that each determinant in this product can ultimately be expressed in terms of the (inverse of the) determinant of $\tilde{\mathcal{H}}_{1,x}$---each $a_i$ depends on $a_1^{-1} = \textrm{adj} \, a_1 \, \textrm{det}^{-1} a_1$---such that it is not possible to find eigenvalue solutions by simply setting each determinant, $\textrm{det}\, a_i$, to zero separately. Instead, one should explicitly reduce this equation by making use of the explicit form of $a_i$ to obtain a polynomial equation of the order of $2N$. Interestingly, as we will see in the following, an elegant and straightforward solution can be derived for some special points in the spectrum.

{\em Degeneracies.} We notice that Eq.~(\ref{eq:rescaling_term}) equals zero, when $h_2 = 0$ and $h_3 = 0$ simultaneously, i.e., when $k_y, k_z \in \{0 , \pi \}$. As a consequence, we straightforwardly find that $a_i = a_1, \, \forall i$, such that
\begin{equation*}
\textrm{det}\, \tilde{\mathcal{H}}_{N,x} = \left(\textrm{det}\, a_1\right)^N = 0, \quad  k_y, k_z \in \{0, \pi \},
\end{equation*}
which, by making use of Eq.~(\ref{eq:det_of_small_block}), leads to
\begin{equation}
\left( i h_0 - \lambda\right)^{2N} = 0.
\end{equation}
We thus find that at these specific values of $k_y$ and $k_z$, all eigenvalues $\lambda$ are $2N$-fold degenerate, and equal to $i h_0$, where
\begin{align*}
\lambda &= i (h + 2), && \textrm{when} \, \, (k_y,k_z) = (0,0), \\
\lambda &= i \, h, && \textrm{when} \,\,  (k_y,k_z) = P[(0,\pi)], \\
\lambda &= i (h - 2), && \textrm{when} \,\,  (k_y,k_z) = (\pi,\pi).
\end{align*}

{\em Eigenvectors at EPs.} Next, we turn to the eigenvectors at these ${\bf k}$ points, i.e.,
\begin{equation*}
\mathcal{H}_{N,x} \Phi = \lambda \Phi, \qquad X \mathcal{H}_{N,x} = \lambda X, \label{eq:eigenvector_eq}
\end{equation*}
where $\Phi = (\phi_{A,1}, \, \phi_{B,1}\ , \ldots, \, \phi_{A, N}, \, \phi_{B,N})^T$ and $X = (\chi_{A, 1}, \, \chi_{B,1}\ , \ldots, \, \chi_{A,N}, \, \chi_{B,N})$ are the right and left eigenvectors, respectively, with the index $A, B$ referring to the degrees of freedom inside the unit cell, as before. Using that $\mathcal{H}_{1,x}$ is an identity matrix with $i h_0 = \lambda$ on the identity for $k_y, k_z \in \{0,\pi\}$, we find, for the right eigenvectors,
\begin{align}
\lambda \phi_{A,1} + \frac{i}{2} \left(\phi_{A,2} + \phi_{B,2}\right) &= \lambda \phi_{A,1}, \label{eq:boundary_one_equality}\\
\lambda \phi_{B,1} + \frac{i}{2} \left(\phi_{A,2} + \phi_{B,2}\right) &= \lambda \phi_{B,1},
\end{align}
and
\begin{align}
\lambda \phi_{A, N} + \frac{i}{2} \left(\phi_{A, N-1} - \phi_{B, N-1}\right) &= \lambda \phi_{A, N}, \\
\lambda \phi_{B,N} + \frac{i}{2} \left(\phi_{B, N-1} - \phi_{A, N-1}\right) &= \lambda \phi_{B,N},
\end{align}
for the two boundaries, and
\begin{align}
\lambda \phi_{A, n} + &\frac{i}{2} \left(\phi_{A, n-1} - \phi_{B, n-1} + \phi_{A, n+1} + \phi_{B,n+1}\right) \nonumber \\
&= \lambda \phi_{A,n}, \\
\lambda \phi_{B,n} + &\frac{i}{2} \left(\phi_{B, n-1} - \phi_{A, n-1} + \phi_{A, n+1} + \phi_{B,n+1} \right) \nonumber \\
&= \lambda \phi_{B, n}, \qquad \forall n \in {2, \ldots, N-1}, \label{eq:bulk_equality}
\end{align}
for the bulk. From these equalities, we obtain the following:
\begin{align*}
&\phi_{A, n-1} - \phi_{B, n-1} = 0, \qquad  & &\phi_{A, n+1} + \phi_{B,n+1} = 0, \\
&\phi_{A,2} + \phi_{B,2} = 0, \qquad & &\phi_{A, N-1} - \phi_{B, N-1} = 0,
\end{align*}
for $n \in {2, \ldots, N-1}$, such that we immediately find
\begin{align*}
\phi_{A, n} &= \phi_{B, n} = 0, \qquad \forall n \in\{2, \ldots, N-1\}, \\
\phi_{A,1} &= \phi_{B,1}, \qquad \phi_{A,N} = -\phi_{B,N}.
\end{align*}
The Hamiltonian matrix $\mathcal{H}_{N,x}$ thus has two linearly independent eigenvector solutions when $k_y ,k_z \in \{0, \pi\}$, namely, $\Phi_{1} = (1 , 1, 0 ,\dots, 0)^T$ and $\Phi_{2} = (0 ,\dots, 0, -1, 1)^T$.

When repeating the same exercise for the left eigenvectors, we find
\begin{align*}
\lambda \chi_{A,1} + \frac{i}{2} \left(\chi_{A,2} - \chi_{B,2}\right) &= \lambda \chi_{A,1}, \\
\lambda \chi_{B,1} + \frac{i}{2} \left(\chi_{B,2} - \chi_{A,2}\right) &= \lambda \chi_{B,1},
\end{align*}
and
\begin{align*}
\lambda \chi_{A,N} + \frac{i}{2} \left(\chi_{A, N-1} + \chi_{B, N-1}\right) &= \lambda \chi_{A, N}, \\
\lambda \chi_{B, N} + \frac{i}{2} \left(\chi_{A, N-1} + \chi_{B, N-1}\right) &= \lambda \chi_{B, N},
\end{align*}
for the two boundaries, and
\begin{align*}
\lambda \chi_{A, n} + &\frac{i}{2} \left(\chi_{A, n-1} + \chi_{B, n-1} + \chi_{A, n+1} - \chi_{B, n+1}\right) \\
&= \lambda \chi_{A, n}, \\
\lambda \chi_{B, n} + &\frac{i}{2} \left(\chi_{A, n-1} + \chi_{B, n-1} - \chi_{A, n+1} + \chi_{B, n+1}\right) \\
&= \lambda \chi_{B, n}, \qquad \forall n \in {2, \ldots, N-1},
\end{align*}
for the bulk. Therefore, following the same steps as before, we arrive at
\begin{align*}
\chi_{A, n} &= \chi_{B, n} = 0, \qquad \forall n \in\{2, \ldots, N-1\}, \\
\chi_{A,1} &= - \chi_{B,1}, \qquad \chi_{A,N} = \chi_{B,N},
\end{align*}
such that we again obtain two linearly independent solutions, i.e., $X_{1} = (1 , -1, 0 ,\dots, 0)^T$, and $X_{2} = (0 ,\dots, 0, 1, 1)^T$. These left eigenvectors are self-orthogonal to the right eigenvectors, i.e., $\braket{X|\Phi} = 0$, and the degenerate eigenvalues with energy $\lambda = i h_0$ thus correspond to exceptional points of the order of $2N-2$.

This piling up of states can also be understood from studying the form of the OBC Hamiltonian at the EPs in more detail, where this argument is inspired by the argument in Sec.~SVIII in Ref.~\onlinecite{Okuma2019}. In the following, we perform a unitary transformation such that $\sigma_x \rightarrow \sigma_z$ in $H_\textrm{NH}({\bf k})$ in Eq.~(\ref{eq:NH_3d_ham}), and find the following Hamiltonian when taking OBC in $x$:
\begin{align*}
H'_{N,x} &= i \, h_0 \sum_{x=1}^N \sum_{j\in\{A,B\}} c^\dagger_{x,j} c_{x,j} \\
& + h_3 \sum_{x = 1}^N \left( c^\dagger_{x,A} c_{x,B}+ c^\dagger_{x,B} c_{x,A} \right) \\
& - i \, h_2 \sum_{x=1}^N \left( c^\dagger_{x,A} c_{x,B} - c^\dagger_{x,B} c_{x,A} \right) \\
& + i \sum_{x = 1}^{N-1}\left( c^\dagger_{x,A} c_{x+1,A} + c^\dagger_{x+1,B} c_{x,B}\right),
\end{align*}
where we use the same conventions for $c^{(\dagger)}_{x,j}$ as before.
First, we note that at the EPs, i.e., $h_2 = h_3 = 0$, this reduces to
\begin{align*}
H'_{N,x, \textrm{EP}} &= \lambda \sum_{x=1}^N \sum_{j\in\{A,B\}} c^\dagger_{x,j} c_{x,j} \\
& + i \sum_{x = 1}^{N-1}\left( c^\dagger_{x,A} c_{x+1,A} + c^\dagger_{x+1,B} c_{x,B}\right),
\end{align*}
where $\lambda$ is the eigenvalue of the EP as before. This Hamiltonian describes two decoupled chains, where there is only hopping to the left in the ``$A$ chain" and only hopping to the right in the ``$B$ chain," thus explaining the appearance of the EPs and the existence of only two eigenvectors. This model thus realizes two decoupled Hatano-Nelson chains \cite{Hatano1996} in the extreme limit, i.e., hopping in one direction only. Second, away from the EPs, $H'_{N,x}$ can be interpreted as describing two coupled Hatano-Nelson chains, which still reside in the extreme limit. Therefore, one would naively expect that when $h_2$ and $h_3$ are small, the eigenstates still pile up albeit now with different eigenvalues. Interestingly, however, both from numerical checks as well as analytical computations (see below), we find that this is not the case. This is in agreement with the result in Ref.~\onlinecite{Okuma2019Topological}, where it is shown that the presence of a magnetic field, which couples the two (pseudo)spin sectors, i.e., coupling $A$ and $B$ to each other, suppresses the NH skin effect. Third, we stress that performing the unitary transformation $(\sigma_x \rightarrow \sigma_z)$ does not alter the physics of our model upon considering OBCs. Indeed, it is possible to show that $U_N = \mathbb{1}_N \otimes U$, where $U^\dagger (\sigma_x, \sigma_y, \sigma_z) U = (\sigma_z, \sigma_y, - \sigma_x)$ with $U = (\sigma_0 - i \sigma_y)/\sqrt{2}$, is a unitary matrix, which transforms $\mathcal{H}_{N,x}$ into $\mathcal{H}'_{N,x}$ according to $U_N^\dagger \mathcal{H}_{N,x} U_N = \mathcal{H}'_{N,x}$. This means that the spectra of $\mathcal{H}_{N,x}$ and $\mathcal{H}'_{N,x}$ are equivalent, and $U^\dagger_N \Phi$ and $XU_N$ are right and left eigenvectors of $\mathcal{H}'_{N,x}$, respectively. Due to its specific form, $U_N$ only acts locally in the eigenvectors and thus does not change their overall behavior. Lastly, as one would expect, we find that $\mathcal{H}'_{N,x}$ equals the Hamiltonian for OBC in $z$, $\mathcal{H}_{N,z}$, with $k_x \rightarrow k_z$ in the latter. Indeed, all results in this paper are independent of whether one takes OBC in $x$, $y$, or $z$ as the Hamiltonian is equivalent in all three directions.

{\em Eigenvectors away from EPs.} Here, we follow the method presented in Ref.~\onlinecite{Yao2018} to find an explicit form for the eigenvectors away from the EPs.
We start by rewriting the equalities for the right eigenvectors in Eqs.~(\ref{eq:boundary_one_equality})--(\ref{eq:bulk_equality}) for general ${\bf k}$, which yields the following:
\begin{align}
(i h_0 + h_3) \phi_{A,1}  - i h_2 \phi_{B,1} + \frac{i}{2} \left(\phi_{A,2} + \phi_{B,2}\right) &= \lambda \phi_{A,1}, \label{eq:boundary_one_generic} \\
i h_2 \phi_{A,1} + (i h_0 - h_3) \phi_{B,1} + \frac{i}{2} \left(\phi_{A,2} + \phi_{B,2}\right) &= \lambda \phi_{B,1},
\end{align}
and
\begin{gather}
(i h_0 + h_3) \phi_{A, N} - i h_2 \phi_{B,N}+ \frac{i}{2} \left(\phi_{A, N-1} - \phi_{B, N-1}\right) \nonumber \\
= \lambda \phi_{A, N}, \\
i h _2 \phi_{A,N} + (i h_0 - h_3) \phi_{B,N} + \frac{i}{2} \left(\phi_{B, N-1} - \phi_{A, N-1}\right) \nonumber \\
= \lambda \phi_{B,N}, \label{eq:boundary_two_generic}
\end{gather}
for the two boundaries, and
\begin{gather}
(i h_0 + h_3) \phi_{A, n} - i h_2 \phi_{B, n}+ \frac{i}{2} \left(\phi_{A, n-1} - \phi_{B, n-1} \right) \nonumber \\
+ \frac{i}{2} \left( \phi_{A, n+1} + \phi_{B,n+1}\right) = \lambda \phi_{A,n}, \label{eq:bulk_generic_one_equality}\\
i h_2 \phi_{A,n} + (i h_0 - h_3) \phi_{B,n} + \frac{i}{2} \left(\phi_{B, n-1} - \phi_{A, n-1} \right) \nonumber \\
 +  \frac{i}{2} \left( \phi_{A, n+1} + \phi_{B,n+1} \right) =\lambda \phi_{B, n}, \qquad \forall n \in {2, \ldots, N-1}, \label{eq:bulk_generic_two_equality}
\end{gather}
for the bulk.

Next, we make the following ansatz for the eigenfunction in unit cell $n$ \cite{Yao2018}:
\begin{equation*}
\Phi_n = \begin{pmatrix}
\phi_{A,n} \\
\phi_{B,n}
\end{pmatrix} = \sum_j r_j^n \begin{pmatrix}
\phi_{A}^{(j)} \\
\phi_{B}^{(j)}
\end{pmatrix}.
\end{equation*}
Plugging this into Eqs.~(\ref{eq:bulk_generic_one_equality}) and (\ref{eq:bulk_generic_two_equality}), we find
\begin{gather}
\left[i h_0 + h_3 - \lambda + \frac{i}{2} \left(r^{-1} + r\right) \right] \phi_{A} = \nonumber \\
 \left[ i h_2 + \frac{i}{2} \left( r^{-1} - r \right) \right] \phi_B, \label{eq:solutions_for_A_in_terms_of_B_one}\\
 \left[i h _2 -  \frac{i}{2} \left(r^{-1} - r \right)  \right] \phi_A = \nonumber \\
 \left[-i h _0 + h_3 + \lambda - \frac{i}{2} \left( r^{-1} +r \right) \right] \phi_B, \label{eq:solutions_for_A_in_terms_of_B_two}
\end{gather}
where we have dropped the index $j$ for brevity, which leads to
\begin{equation*}
\left(h_0 + i \lambda \right) \left(r^{-1} + r\right)  = - 1 -  \left(h_0 + i \lambda \right)^2 - h_2^2 - h_3^2.
\end{equation*}
Rewriting this equation, we find the following second-order polynomial:
\begin{equation*}
r^2 \left(h_0 + i \lambda \right) + r \left[1 +  \left(h_0 + i \lambda \right)^2 + h_2^2 + h_3^2\right] + \left(h_0 + i \lambda \right)  = 0
\end{equation*}
with solutions
\begin{equation*}
r_\pm =  \frac{-1 - \left(h_0 + i \lambda \right)^2 - h_2^2 - h_3^2 \pm \sqrt{D}}{2 \left(h_0 + i \lambda \right)},
\end{equation*} \begin{equation*}
D = \left[1 +  \left(h_0 + i \lambda \right)^2 + h_2^2 + h_3^2\right]^2 - 4 \left(h_0 + i \lambda \right)^2.
\end{equation*}
Therefore, we find that the general solution for $\Phi_n$ reads
\begin{equation}
\Phi_n = \begin{pmatrix}
\phi_{A,n} \\
\phi_{B,n}
\end{pmatrix} = r_+^n \begin{pmatrix}
\phi_{A}^{(+)} \\
\phi_{B}^{(+)}
\end{pmatrix} + r_-^n \begin{pmatrix}
\phi_{A}^{(-)} \\
\phi_{B}^{(-)}
\end{pmatrix}, \label{eq:ansatz_for_eigenfct_sol}
\end{equation}
with $r_\pm$ as specified above. Note that $r_+ \rightarrow 1$ when $i h_0 \rightarrow \lambda$, i.e., at the EPs, whereas $r_-$ diverges. However, as we are interested in the eigenvectors away from the the EPs, i.e., $\lambda \neq i h_0$, this is not a problem.

From the explicit expressions of $r_\pm$, we immediately find that $r_+ r_- = 1$.
Moreover, we know from Ref.~\onlinecite{Yao2018} and further generalizations in Ref.~\onlinecite{Yokomizo2019} that $|r_+| = |r_-|$ for the continuum eigenfunctions, such that we trivially find $|r_i| = 1$.
This means that the bulk states do not pile up as the localization coefficients of the bulk states $r_i$ equal a phase, such that the skin effect does not appear away from the EPs.

\end{document}